\documentclass{article}
\usepackage{arxiv}
\usepackage[utf8]{inputenc} 
\usepackage[T1]{fontenc}    
\usepackage{hyperref}       
\usepackage{url}            
\usepackage{booktabs}       
\usepackage{amsfonts}       
\usepackage{nicefrac}       
\usepackage{microtype}      
\usepackage{lipsum}		
\usepackage{graphicx}
\usepackage{etoolbox}
\apptocmd{\sloppy}{\hbadness 10000\relax}{}{}

\usepackage[inline]{enumitem}
\usepackage{xcolor}
\usepackage{amsmath,amssymb}
\usepackage[outdir=./]{epstopdf}

\title{Disrupting Resilient Criminal Networks through Data Analysis: The case of Sicilian Mafia}

\author{ \href{https://orcid.org/0000-0002-2367-6084}{\includegraphics[scale=0.06]{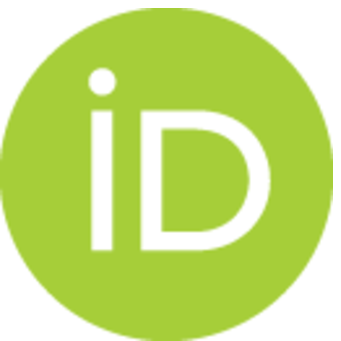}\hspace{1mm}Lucia Cavallaro}\thanks{Corresponding author. Alternative e-mail address \texttt{lucia.cavallaro3@gmail.com}} \\
	Department of Engineering and Technology\\
	University of Derby\\
	Derby, DE22 1GB, UK \\
	\texttt{L.Cavallaro@derby.ac.uk} \\
	\And
	\href{https://orcid.org/0000-0001-9517-4131}{\includegraphics[scale=0.06]{orcid.eps}\hspace{1mm}Annamaria Ficara} \\
	Department of Mathematics and Computer Science\\
	University of Palermo\\
	Palermo, 90123, Italy \\
	\texttt{aficara@unime.it} \\
	\And
	\href{https://orcid.org/0000-0001-7421-216X}{\includegraphics[scale=0.06]{orcid.eps}\hspace{1mm}Pasquale De Meo} \\
	{Department of Ancient and Modern Civilizations}\\
	University of Messina\\
	Messina, 98122, Italy \\
	\texttt{pdemeo@unime.it} \\
	\And
	\href{https://orcid.org/0000-0003-1528-7203}{\includegraphics[scale=0.06]{orcid.eps}\hspace{1mm}Giacomo Fiumara} \\
	MIFT Department\\
	University of Messina\\
	Messina, 98166, Italy \\
	\texttt{gfiumara@unime.it} \\
	\And
	\href{https://orcid.org/0000-0002-0369-8235}{\includegraphics[scale=0.06]{orcid.eps}\hspace{1mm}Salvatore Catanese} \\
	MIFT Department\\
	University of Messina\\
	Messina, 98166, Italy \\
	\texttt{salvocatanese@gmail.com} \\
	\And
	\href{https://orcid.org/0000-0003-4193-9842}{\includegraphics[scale=0.06]{orcid.eps}\hspace{1mm}Ovidiu Bagdasar} \\
	Department of Engineering and Technology\\
	University of Derby\\
	Derby, DE22 1GB, UK \\
	\texttt{O.Bagdasar@derby.ac.uk} \\
	\And
	\href{https://orcid.org/0000-0002-2773-4421}{\includegraphics[scale=0.06]{orcid.eps}\hspace{1mm}Antonio Liotta} \\
	School of Computing\\
	Edinburgh Napier University\\
	Edinburgh, EH10 5DT, UK \\
	\texttt{A.Liotta@napier.ac.uk} \\
}

\begin{document}
\maketitle
\raggedbottom
\begin{abstract}
Compared to other types of social networks, criminal networks present hard challenges, due to their strong resilience to disruption, which poses severe hurdles to law-enforcement agencies. Herein, we borrow methods and tools from Social Network Analysis to 
\begin{enumerate*}[label=(\roman*)] 
\item unveil the structure of Sicilian Mafia gangs, based on two real-world datasets, and  
\item gain insights as to how to efficiently disrupt them. 
\end{enumerate*}
Mafia networks have peculiar features, due to the links distribution and strength, which makes them very different from other social networks, and extremely robust to exogenous perturbations. Analysts are also faced with the difficulty in collecting reliable datasets that accurately describe the gangs' internal structure and their relationships with the external world, which is why earlier studies are largely qualitative, elusive and incomplete. An added value of our work is the generation of two real-world datasets, based on raw data derived from juridical acts, relating to a Mafia organization that operated in Sicily during the first decade of 2000s. We created two different networks, capturing phone calls and physical meetings, respectively. Our network disruption analysis simulated different intervention procedures: 
\begin{enumerate*}[label=(\roman*)] 
\item arresting one criminal at a time (sequential node removal); and 
\item police raids (node block removal). 
\end{enumerate*}
We measured the effectiveness of each approach through a number of network centrality metrics. We found Betweeness Centrality to be the most effective metric, showing how, by neutralizing only the 5\% of the affiliates, network connectivity dropped by 70\%. We also identified that, due the peculiar type of interactions in criminal networks (namely, the distribution of the interactions frequency) no significant differences exist between weighted and unweighted network analysis. Our work has significant practical applications for tackling criminal and terrorist networks.
\end{abstract}

\keywords{Criminal Networks \and Complex Networks \and Social Network Analysis \and Graph Theory}

\section{Introduction}
\label{sec:introduction}
The Sicilian Mafia is a very specific type of criminal organization, which originated in Sicily and has now spread worldwide~\cite{franchetti1925sicilia, doi:10.1111/j.1745-9133.2005.00306.x, Mastrobuoni2012}, taking control of entire economic sectors and influencing the social and political life of entire countries. Compared to other criminal organizations, Mafia has a unique \textit{modus operandi}: it appears as a collection of loosely coupled groups (referred to as {\em cosche}, {\em gangs}, or {\em families}), which last for several generations. Members of a Mafia gang are tied by strong links, thus making the organization efficient and capable of swiftly re-organizing itself to pursue the most profitable activities and adjust to law enforcement operations. In fact, Mafia tends to create deep roots into the very fabric of society, to the point that it becomes ``impossible to destroy without a radical change in social institutions" (in the words of Italian politician Leopoldo Franchetti, 1876~\cite{franchetti1925sicilia}).

Given the social embeddedness of organized crime and, in particular, of Mafia-like organizations, the analysis of the social structure of Sicilian Mafia syndicates has generated significant scientific interest~\cite{doi:10.1177/1477370807084225,66338547350542128865ef8290c2d860}. 
Herein, we borrow methods and tools from Social Network Analysis (SNA) to
\begin{enumerate*}[label=(\roman*)] 
\item unveil the structure and organization of Sicilian Mafia gangs, based on two real-world datasets, and 
\item gain insights as to how to efficiently disrupt them.
\end{enumerate*}
SNA is typically employed by Law Enforcement Agencies (LEA) to analyze criminal networks, investigate the relations among criminals, and evaluate the effectiveness of law enforcement interventions aimed at disrupting criminal networks~\cite{Duijn2014}.
Morselli~\cite{Morselli2003} studied the connections within the Gambino, New York based family, focusing on the career of one of its members, Saul Gravano. 
McGloin~\cite{doi:10.1111/j.1745-9133.2005.00306.x} analyzed the network structure of street gangs in Newark, New Jersey.
Calderoni~\cite{Calderoni2012} showed that high-status Mafia members were able to indirectly manage illicit drug traffics, by keeping the middle-level criminals in more central and visible positions. 

On the other hand, strategies for criminal network disruption can be divided into two main approaches~\cite{Schwartz2009}: the {\em human capital} and the {\em social capital} approach.
The former originates from economics, and refers to the personal attributes and/or resources possessed by actors within a social network. 
Sparrow~\cite{SPARROW1991251} suggested that identifying the individuals who possess many resources and skills offers a great opportunity to damage the criminal network. 
Cornish~\cite{Cornish94theprocedural} introduced the notion of {\em script}, which is borrowed from cognitive science. A script approach is a way to better understand how crimes are committed and how to prevent them. The central element of this approach, the crime script, is a step-by-step account of the actions and decisions involved in a crime. For example, a robbery script can be constructed as a schematic plan of a subway mugging. If the script is correctly identified, it can be used to prevent or disrupt crime commission. 
Later on, Bruinsma and Bernasco~\cite{Bruinsma2004} combined this script concept with SNA, to identify human capital within criminal networks. Morselli and Roy~\cite{doi:10.1111/j.1745-9125.2008.00103.x} applied this method to study ``brokerage roles'' within criminal. 

By contrast, the social capital network-disruption strategy ~\cite{coleman1990foundations, vanderHulst2009} refers to the connections or ties between actors in a network. It is through these connections that actors can have strategic positions, exchanging and sharing resources with other actors in the network~\cite{SPARROW1991251, Klerks2001, Natarajan2006, Carley2001, Schwartz2009}. Research in this field is often based on SNA to identify central actors in the network, that are associated with influential or powerful positions of social capital~\cite{Lin2001}. 
There is empirical evidence that {\em brokers} (i.e., the individuals acting as bridges between disconnected subgroups) play important roles in connecting criminal networks, often connecting separate criminal collectives within illegal markets~\cite{Natarajan2006, Calderoni2010, doi:10.1111/j.1745-9125.2008.00103.x, Morselli2001, Campana2012,FERRARA20145733, Ferrara2014,AGRESTE201630}. 

Through SNA, a number of interesting findings have been derived over the years. Agreste {\em et al.}~\cite{AGRESTE201630} applied percolation theory to efficiently dismantle mafia syndicates.
Peterson~\cite{peterson1994applications} argued that the most central actors in covert networks might also be the most visible and for this reason the most likely to be detected. 
Spapens~\cite{Spapens2010} identified a brokerage role within Dutch ecstasy production value chains, observing that brokers not only increase ``social capital'' within these criminal collectives, but also add ``human capital''. Bright et al.~\cite{doi:10.1080/17440572.2017.1377614} investigated the effectiveness of five law enforcement interventions in disrupting and dismantling criminal networks, using both the social and human capital approaches in criminal networks. They showed how the removal of actors based on the  Betweenness centrality metric was the most efficient strategy.

In this paper we focus particularly on one aspect of {\em network resilience} that is the ability of criminal networks to survive to the actions of LEAs. However, network resilience also relates to getting networks reorganized after perturbations (e.g., police raids) to reestablish connectivity. This latter aspect has not been investigated herein because our networks are treated as a static dataset. Nonetheless, this work is a significantly extended version of the early proof-of-concept results published in our recent conference paper~\cite{Catanese2016}.

Several authors~\cite{Bouchard2007, AYLING2009182} have described the concept of {\em network resilience} considering two main aspects: 
\begin{enumerate*}[label=(\roman*)]
\item the capacity to absorb and thus resist disruption,
\item the capacity to modify the network internal structure and strategies to adapt to external pressure.
\end{enumerate*}
Resilience depends on the level of {\em redundancy} present in the criminal network. Redundancy reflects the diversity of relationships among the network actors, and is associated with strong connections between these actors~\cite{Williams2001}. Even if some connections are broken, the diversity of different ties between actors allows the network to continue to function. Strong criminal ties offer reciprocated trust in an uncertain and hostile criminal environment~\cite{Calderoni2010}. Replacements with a reliable reputation are therefore often found within the social connections directly embedding the actors involved within the criminal business process. These criminal connections often initiate from already established social networks of kinship, friendship or affective ties~\cite{10.2307/3006013, doi:10.1177/1477370807084225}. This means that replacements are often found within short social distances.

We have already mentioned that Sicilian Mafia is very different from other criminal organizations. Mafia networks have peculiar features, due to the links distribution and strength, which makes them extremely robust to exogenous perturbations. Analysts are also faced with the difficulty in collecting reliable datasets that accurately describe the gangs' internal structure and their relationships with the external world, which is why earlier studies are largely qualitative, elusive and incomplete. An added value of our work is the generation of two real-world datasets, which we have anonymized (to eliminate sensitive data) and made publicly available online at: \url{https://github.com/lcucav/networkdistruption}. These are based on raw data derived from juridical acts, relating to a Mafia gang that operated in Sicily (Italy) during the first decade of 2000s. We created two very different networks, capturing phone calls and physical meetings, respectively, which we have characterized in our recent conference paper~\cite{10.1007/978-3-030-36683-4_36}. 

Our datasets relate to a Mafia syndicate acting as a link between prominent criminal families, operating in the two biggest cities (Palermo and Catania). The {\em phone calls} dataset has been derived from eavesdropping, while the {\em meetings} dataset has been derived from police surveillance data. Both datasets are unidirected networks. For each one, we have created a {\em weighted} graph version (considering the frequency of interactions between individuals, or nodes), as well as an {\em unweighted} version (accounting only for connections).

The present study goes well beyond the initial characterization of these datasets~\cite{10.1007/978-3-030-36683-4_36}, investigating network robustness across different scenarios, pinpointing the most effective metric, and demonstrating an effective network disruption strategy. We simulate two types of police operations: 
\begin{enumerate*}[label=(\roman*)] 
\item arresting one criminal at a time (sequential node removal), and 
\item police raids (node block removal). 
\end{enumerate*}
We evaluate how the different types of networks are impacted by these two types of {\em perturbations}, in terms of structure and connectivity. 

Generally, the resilience of a network is the result of several factors, deriving from the network structure, the position of the nodes, and the human capabilities. The latter refers to personal skills and competences (e.g., pharmacological and chemical knowledge are required in synthetic drug synthesis processes). Thus, we base our attack strategy on the {\em social capital} approach. We employ SNA methods to identify the actors having a high level of social capital. These are typically the most influential individuals, with a {\em central} role in the criminal network. To this end, we put to test four different centrality metrics, namely: 
\begin{enumerate*}[label=(\roman*)]
\item Degree centrality,
\item Betwenness centrality,
\item Katz centrality, and 
\item  Collective Influence. 
\end{enumerate*}
It is worth recalling from SNA that the {\em  Degree centrality} helps identifying network hubs (i.e., the focal points). However, contrary to other types of networks, in criminal networks communications are not necessarily mediated via hubs, which are more visible and, thus, more vulnerable. On the other hand, by weighing the communication paths (rather nodes in isolation), {\em betweeness centrality} pinpoints those nodes that play an important role in multiple communication paths. We have therefore hypothesized that betweeness centrality could help removing the individuals that are crucial in maintaining the information network. In turn, removing those individuals would maximize network disruption, which is our aim. 

For the sake of completeness, we also considered two more prominent centrality metrics. {\em Katz} computes the relative influence of a node, measuring the number of its immediate neighbours (first degree nodes) and also all other nodes in the network that connect to the node itself through these immediate neighbours. Finally, {\em Collective Influence} establishes the centrality of a node in a criminal network taking into account the degree of the node's neighbours at a given distance $l$ from it.

Our analysis, as detailed in the following, Betweeness Centrality to be the most effective metric, showing how, by neutralizing only 5\% of the affiliates, network connectivity dropped by 70\%. We also identified that, due the peculiar type of interactions in criminal networks (namely, the distribution of the interactions frequency) no significant differences exist between weighted and unweighted network analysis. Our work has significant practical applications for tackling criminal and terrorist networks.

\section*{Materials and Methods}
\label{sec:methods}
\subsection*{Dataset Description}
\label{subsec:dataset}

\begin{figure}[!ht]
\centering
\includegraphics[scale=0.32]{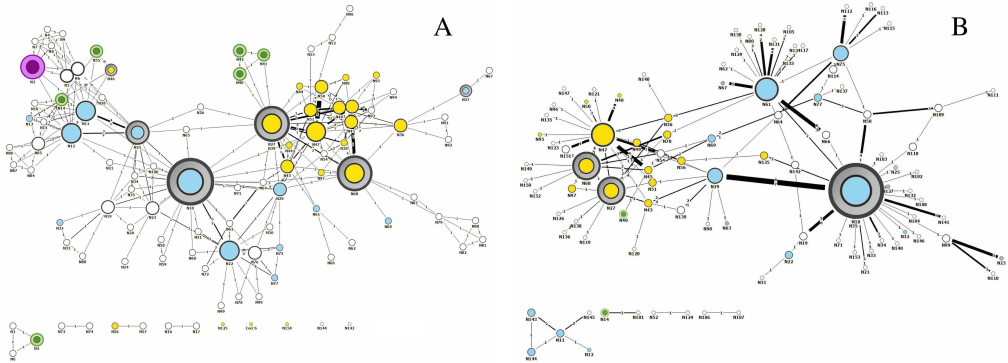}
\caption{{\bf Dataset visualization. A: {\em Meetings} network. B: {\em Phone Calls} network.} The colors represent different clans, including: the ``Mistretta'' family (turquoise); the ``Batanesi'' clan (yellow); other Mafia families (green and purple). The {\em double circled} nodes depict the {\em bosses}, i.e., those leaders who have been investigated for having promoted, organized, or directed the Mafia association. Finally, the white nodes represent other subjects who are close to a family, but are not classifiable in any of the previous categories (e.g., entrepreneurs who take part to illegal activities, but do not necessarily belong to one of the Mafia families). In both graphs, the edges' width is proportional to the number of meetings or phone calls, while the size of the nodes is proportional to their degree. The nodes shown at the bottom are isolated from the largest connected component.}
\label{fig:graphs}
\end{figure}

In this section we explain how we have created the datasets (available at: \url{https://github.com/lcucav/networkdistruption}), generating two criminal networks directly from Court data. These are represented in Fig.~\ref{fig:graphs} as undirected and weighted graphs, whereby nodes represent the individuals and links give an account of interactions among them. The {\em Meetings} dataset, accounts for the physical meetings among criminals; the {\em Phone Calls} dataset refers to phone calls among individuals.

Our datasets were derived from the pre-trial detention order, issued by the Court of Messina’s preliminary investigation judge on March 14, 2007, which was towards the end of the major anti-mafia operation referred to as ``Montagna Operation''. This operation was concluded in 2007 by the Public Prosecutor’s Office of Messina (Sicily) and was conducted by the Special Operations Unit of the Italian Police (the R.O.S. {\em Reparto Operativo Speciale} of the {\em Carabinieri} is specializing in anti-Mafia investigations). 
This particular investigation was a prominent operation focused on two Mafia clans, known as the ``Mistretta'' family and the ``Batanesi clan''. From 2003 to 2007, these families were found to had infiltrated several economic activities including major infrastructure works, through a cartel of entrepreneurs close to the Sicilian Mafia. 

The main characteristics of our datasets are summarized in Table~\ref{table:networks}, including number of nodes and edges, the maximum weight, the maximum frequency in the affiliates’ interactions, the average degree (neighbours per node), and the highest number of individuals required to connect mobsters (considering all the shortest paths on both datasets). It is worth noticing that the two datasets have 47 nodes in common.

\begin{table}
\caption{{\bf Characteristics of {\em Meetings} and {\em Phone Calls} networks.}}
\centering
\begin{tabular}{|l|c|c|}
\hline
	Parameter     & Meeting     & Phone Calls \\ \hline
	No. Nodes & 101 &  100\\ \hline
	No. Edges & 256  & 124\\ \hline
	Max. Weight & 10 & 8\\ \hline
	Avg. Degree & 5.07 & 2.48\\ \hline
	Max. Shortest Path & 7 & 14\\ \hline
	Common nodes & \multicolumn{2}{c|}{47} \\ \hline
\end{tabular}
\label{table:networks}
\end{table} 

Indeed, both networks can be treated as either {\em unweighted} and {\em weighted} networks since, for each pair of individuals, we recorded a coefficient (i.e., the weight $w$) representing the number of times the pair had a meeting (as reported by the police surveillance logs), and the number of times two individuals called each other (reported in police interceptions logs). In SNA terms, these coefficients are known as the strength of the tie binding two individuals.

Finally, the nodes may below to different categories. They are either ``bosses" (i.e., the leaders of the criminal organization), or ``picciotti" (i.e., the soldiers of the organization). Nonetheless, there are also roles that are not necessarily pointing to criminals. For instance the fruit seller or the baker may somehow be connected to members of the organization for a range of reasons.

\subsection*{Background}
\label{subsec:background}

In this paper we consider two weighted and undirected criminal networks or graphs. For convenience, we first summarise some key concepts of network theory \cite{barabasi2016network} and formalize the problem at hand.

An {\em unweighted graph} is defined as a graph $G = (V, E)$, where $V$ is a set of vertices (also called nodes, actors) and $E \subseteq V \times V$ is a set of edges (also called links or ties). 

A {\em weighted graph} $G = (V, E, W)$ is defined as a triplet consisting of a finite set of $n$ nodes or vertices $V$, a set of edges $E \subseteq V \times V$, and a set of positive weights $W: E \rightarrow R_{++}$ defined on each edge.

When all the edges are bidirectional, the graph is called {\em undirected}. 

A {\em path} from nodes $u$ to $v$ is a sequence of nodes starting with $u$ and ending with $v$, such that between consecutive nodes there is a link. The {\em path cost} is the sum of the weights of all links along the path. 
A {\em shortest path} from $u$ to $v$ is a path along which the path cost is minimal (there may exist multiple shortest paths). In weighted graphs, it equals the sum of the weights along the path. For unweighted graphs, a shortest path is that (or those) consisting of the fewest number of links. 

A {\em connected component} \cite{10.2307/4615713} of an undirected graph is a subgraph in which any two vertices are connected to each other by paths, and which is connected to no additional vertices outside of this component (i.e., the connected components represent a partition of the whole graph, sometimes called {\em supergraph}). The {\em largest connected component} (LCC) is the biggest one among all the connected components in a graph.

A commonly used representation of a graph is the {\em adjacency matrix} \cite{BUSATO2017163}. This is a $V \times V$ matrix $A = [f(i,j)]$ where each element $f(i,j)$ contains the attributes of the edge $(i,j)$. If the edges have no specific attribute, then the graph can be represented by a boolean matrix.
If the graph is weighted, the matrix is defined as follows:
\begin{eqnarray}
\label{eq:adjm}
\mathrm{A[i,j]} = \begin{cases} \begin{array}{l}
     0 \hspace{1.55cm} \textrm{ if } \enspace i = j\\
     w(i,j) \qquad \textrm{ if } \enspace i \neq j 
     \enspace \textrm{ and } \enspace (i,j) \in E\\
    \infty \hspace{1.30cm} \textrm{ if }\enspace i \neq j \enspace \textrm{  and } \enspace (i,j) \not\in E
\end{array}
\end{cases}
\end{eqnarray}

\textit{Centrality} is a key concept in network analysis, and refers to the importance of a node in a network.
There are multiple measures for network centrality in use within SNA, but the two most common centrality measures relating to strategic positions are Degree centrality and Betweenness centrality~\cite{SPARROW1991251, Klerks2001}.

{\it Degree centrality}~\cite{FREEMAN1978215} is a measure which evaluates the local importance of a node within the graph; given a node $i$, the  Degree centrality $C_D(i)$ of $i$ is defined by:
\begin{eqnarray}
\label{eq:deg}
\mathrm{C_D(i)} = \sum\limits_{j=1, j\neq i}^{n} a_{ij},
\end{eqnarray}
where each $j=1,\dots,n$ is a node, $n$ is the total number of nodes, and $a$ is the adjacency matrix, in which $a_{ij} = 1$ if there is an edge from $i$ to $j$, and $0$ otherwise.

The {\em node degree} represents the number of edges adjacent to the node, while the {\em weighted node degree} is the sum of the edge weights, for edges incident to that node. This measure has been formalized as follows~\cite{Barrat2004}:
\begin{eqnarray}
\label{eq:wdeg}
\mathrm{C^W_D(i)} = \sum\limits_{j=1, (i,j)\in E}^{n} w_{ij},
\end{eqnarray}
where $w$ is the weighted adjacency matrix, in which $w_{ij}$ is greater than $0$ if the node $i$ is connected to node $j$, and the value represents the weight of the edge.

Some nodes may play an important role in propagating information because they act as bridges between separate regions of a graph and so they have the potential to slow down (or magnify) the information flow from one region to another. Such nodes are said to have a high value of {\em  Betweenness centrality}~\cite{Brandes2008OnVO}. 
Specifically, the (shortest-path) betweenness $C_B(i)$ of a node $i$ is defined as follows:
\begin{eqnarray}
\label{eq:bet}
\mathrm{C_B(i)} = \sum\limits_{s,t \in N} \frac{\sigma (s,t | i)}{\sigma (s,t)},
\end{eqnarray}
where $ \sigma (s, t)$ is the number of shortest paths between an arbitrary pair of nodes $s$ and $t$, while $ \sigma (s, t | i)$ denotes those shortest paths passing through the node $i$.

When computing the Betweenness centrality for both unweighted and weighted networks, the formula must account instead for shortest paths. For the algorithm we use breadth-first search (BFS) for unweighted and Dijkstra’s algorithm for weighted graphs. 

{\em Katz centrality}~\cite{Katz1953} in another centrality measure, which defines the centrality for a node based on the centrality of its neighbours. It is a generalization of the eigenvector centrality, and for a node $i$ it is defined as follows:
\begin{eqnarray}
\label{eq:kats}
\mathrm{x_i} = \alpha \sum_{j} A_{ij} x_j + \beta,
\end{eqnarray}
where $A$ is the adjacency matrix of the graph $G$, whose eigenvalues are denoted by $\lambda_i$, $i=1,\dots,n$.
The parameter $\beta$ controls the initial centrality and satisfies: 
$$\alpha < \frac{1}{
\max\{\lambda_{i}: \: 1\leq i \leq n\}}.$$

The Katz centrality for weighted networks can be computed in a similar way, but in this case we have to use the weighted adjacency matrix instead. 

Another useful network metric is the {\em Collective Influence} (CI)~\cite{Morone2015}, which computes the centrality or influence of a node in a network according to the formula:
\begin{eqnarray}
\label{eq:ci}
\mathrm{CI_{\ell}(i)} = (k_i - 1) \sum_{j \in \delta B(i, \ell)} (k_j - 1),
\end{eqnarray}
where $k_i$ is the degree of node $i$, $B(i, \ell)$ is the ball of radius $\ell$ centered on node $i$, and $\delta B(i, \ell)$ is the frontier of the ball, that is, the set of nodes at distance $\ell$ from $i$ (the distance between two nodes is defined as the number of edges of the shortest path connecting them). To compute $CI_{\ell}(i)$, we first find the nodes on the frontier $\delta B(i, \ell)$. 
To compute the CI in a weighted network, we have to substitute the degree $k$ of a node by his weighted degree given by formula \eqref{eq:wdeg}.

\subsection*{Experimental method}
This section briefly describes the experimental process employed to disrupt the two networks and evaluate the effects of node removal, under different conditions and strategies. Both datasets were used, under both unweighted and weighted conditions. Two node removal strategies have been studied. These are iterative procedures in which the nodes have been removed in decreasing order of their centrality score. After the node removal stage, the size of the LCC is updated and the process resumes. 

We define $LCC_i(G_i)$ as the Largest Connected Component size at the $i$th iteration of the algorithm. Thus, $LCC_0(G_0)$ denotes the LCC size in the unperturbed graph $G_0$ before the node removal process. We, then, define $\rho\in[0,1]$ as follows:
\begin{equation}
        \label{eq:fluctuation}
        \rho_i = 1 - \Big|\frac{LCC_{i}(G_i) - LCC_0(G_0)}{LCC_{0}(G_0)}\Big|
\end{equation}{}
Note that $\rho_0 = 1$ and $\rho_n = 0 $, whereby $n$ represents the last iteration.

Both strategies may be summarized as follow: 
\begin{enumerate}
    \item The Largest Connected Component size, $LCC_0$ of the initial graph, $G_0$ is computed.
    \item Depending on the removal strategy. Either the highest-rank node (in the sequential strategy) or the set of the five highest-rank nodes (in the block strategy) are removed. Ranks are computed with the current centrality score. The new graph $G_1$ is obtained.
    \item $LCC_1(G_1)$ is computed, and $\rho_1$ is now defined. 
    \item Depending on the removal strategy. Either the second highest-rank node (in the sequential strategy) or the second set of five more influential nodes (in the block strategy) are removed from $G_1$, based on the current centrality metric. The new graph $G_2$ is obtained.
    \item $LCC_2(G_2)$ is computed, and $\rho_2$ is now defined. 
    \item The Steps from the $2^{nd}$ to the $5^{th}$ are repeated until the graph size $G_n$ is equal to zero.
\end{enumerate}{}

\paragraph{Sequential Nodes Removal.} It simulates the scenario in which affiliates are arrested one-by-one by the police. 

\paragraph{Block Nodes Removal.} It simulates the scenario in which affiliates are arrested during a raid by the police. This strategy is similar to the sequential one, with the main difference being that nodes are removed in blocks of five. This block size was found to be adequate to the type and scale of the datasets.

\section*{Results}
\label{results} 
\begin{figure}[!ht]
\centering
\includegraphics[width=0.6\textwidth]{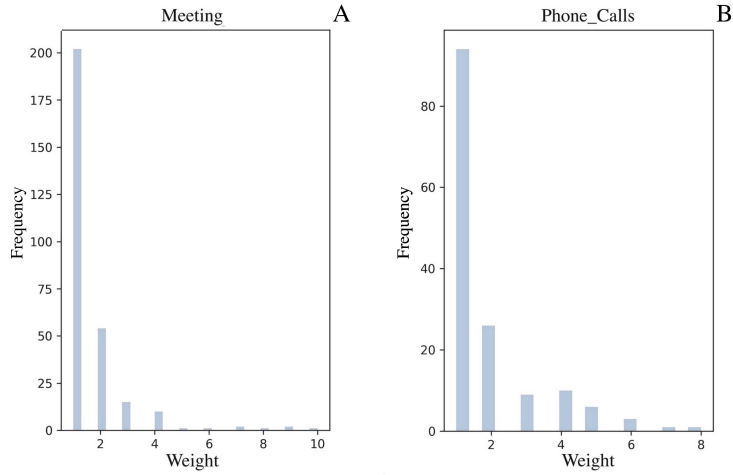}
\caption{{\bf Weights Distribution.} A: \textit{Meeting} Dataset. B: \textit{Phone Calls} Dataset. }
\label{fig:weight_distr}
\end{figure}

In this section the results obtained from our network disruption experiments are shown. We start from the datasets introduced and characterized in our earlier conference paper ~\cite{10.1007/978-3-030-36683-4_36}, in which we defined the weight distributions (Fig.~\ref{fig:weight_distr}) and we conducted the shortest path length analysis. The network disruption analysis is reported next, considering weighted and unweighted graphs scenarios.

\begin{figure}[!ht]
\centering

\includegraphics[width=0.6\textwidth]{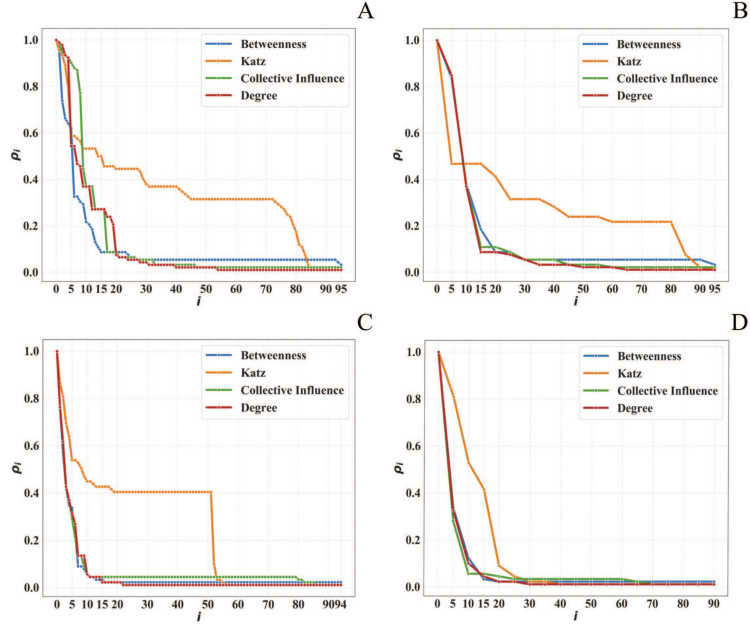}
\centering
\caption{{\bf Weighted networks.} 
A:~\textit{Meeting} dataset, sequential node removal strategy. B:~\textit{Meeting} dataset,  block node removal strategy.
C:~\textit{Phone Calls} dataset, sequential node removal strategy. D:~\textit{Phone Calls} dataset, block node removal strategy.}
\label{fig:weighted}
\end{figure}

\paragraph{Weighted Graphs.}
Fig.~\ref{fig:weighted} shows the results obtained for the cases of sequential and block node removal, for both datasets, and including all four centrality metrics. Remarkably, the Katz coefficient (tuned to the default values of $\alpha=0.1$ and $\beta=1.0$) is the least effective one (i.e, the slowest one) at disrupting the network, in all eight cases, that are: two datsets (Meeting and Phone Calls), two strategies (sequential and block) and two graph structures (weighted and unweighted). To understand this result intuitively, we need to look at the way this centrality metric operates. Katz determines the importance of each node based on the number of {\em walks} that pass through it; but it does not consider their length. Furthermore, shortest path are not considered. In other words, it is allowed for a {\em walk} to visit the same node multiple times. Yet, this is in contrast to how criminals would operate in practice. Affiliates typically prefer to spread the information through a number of intermediaries, to minimize the risk of interception from non-family individuals. This is also coherent with our earlier findings ~\cite{10.1007/978-3-030-36683-4_36}. Ultimately, it would not make sense (and would be unwise) to send the same message multiple times through the same path, would is what Katz would help identifying. Therefore, removing nodes by highest Katz score would not be a winning strategy. 

All the other metrics act better than Katz centrality, and similarly among each other. This happens because of the weights distribution shape (Fig.~\ref{fig:weight_distr}), which exhibits a long tail of nodes, with just a few dominating ones. Thus, after the removal of most central nodes (i.e., the first five iterations), the network gets almost totally disconnected and the remaining nodes have the same weight ($w=1$). Hence, all the metrics focused on either degree (i.e., Degree and Collective Influence), or shortest paths (i.e., Betweenness) follow the same $\rho$ drop speed. On the other hand, Katz centrality with its default parameters focuses on walks of undefined lengths, thus producing a slower $\rho$ drop.

\paragraph{Sequential vs Block Removal.}
Looking at Fig.~\ref{fig:weighted}, with the exception of Katz, no significant differences are visible between the two node-removal strategies (i.e., sequential and block). This is somewhat counter-intuitive, since in real life police raids are typically aimed at breaking up the network more effectively. In our case, this result originates from the particular type of datasets at hand. When we constructed the datasets, we had did not have access to information relating to the way criminals reconstructed their communication channels following arrests. Hence, our network is static (i.e. it misses the network reconfiguration data), which is why our analysis is not fully capturing the dynamic aspects that differentiate sequential and block strategies. 

\paragraph{Weighted vs Unweighted.}

Considering now the differences between weighted and unweighted graph analysis, we notice that the majority of cases do not pinpoint major differences. This was due to the peculiar way in which weights are distributed in criminal networks (as noted in the \textit{Weighted Graphs} paragraph). 

\begin{figure}[!ht]
\centering
\includegraphics[width=0.7\textwidth]{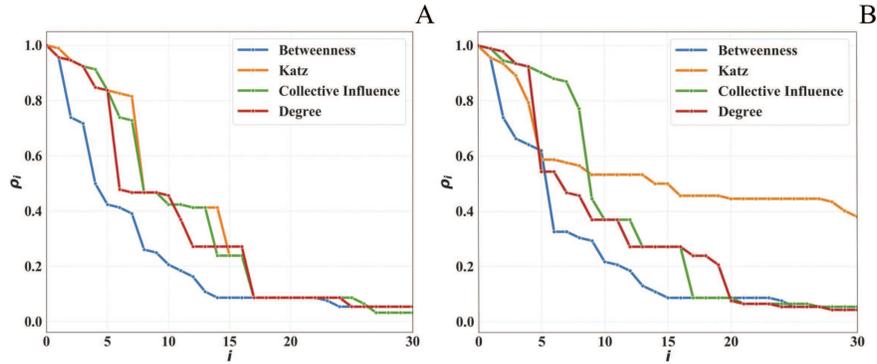}

\caption{{\bf First 30 iterations of the sequential node removal strategy, \textit{Meeting} dataset.} A:~Unweighted Graph. B:~Weighted Graph.}
\label{fig:w_vs_un}
\end{figure}

Nevertheless, interesting differences are visible in the Meeting dataset - sequential node removal (Fig.~\ref{fig:w_vs_un}). The unweighted case is mostly faster than (although occasionally equivalent to) the weighted case. This is because the weights (i.e., the affiliates' interaction frequency) are concentrated in very few individuals (Fig.~\ref{fig:weight_distr}), with most other weights having $w=1$. This is also why, with the exception of the initial transient period (involving very few interactions), most algorithms converge to similar values.

\paragraph{The best centrality metric.}
Comparing the algorithms in  Fig.~\ref{fig:w_vs_un}, it emerges that Betweeness centrality is by far the most effective metric in terms of ability to disrupt the information flow among affiliates.
This is also coherent with literature reports upon criminal networks' SNA, as referred to in the Introduction.

Intuitively, Betweenness centrality outperforms the other metrics, thanks to its operation on paths, rather than on individual nodes degree. This is particularly effective in criminal networks that are devised in such a way as to minimize the path length followed by the information flows, in order to reduce the risk of police interceptions. In fact, Betweeness centrality compromises the most influential paths, leading to a faster drop in $\rho$. At the same time, this feature is what makes Betweeness somehow opposite to Katz centrality, whose goal is to explore walks.

The second worst performer (following Katz, as discussed in \textit{Weighted Graphs}) was found to be ``Collective Influence". This is, again, due to its emphasis on node degree, rather than path length. 
Collective Influence is also showing some differences between the weighted and the unweighted processes, exhibiting a lower $\rho$ drop in the weighed graph. 

A possible explanation is that the weighted case identifies as influential nodes not only those with higher weights on the incident links, but also the nodes having high-weight only on immediate neighbours. This could reflect a typical situation in criminal networks, whereby the top-leaders avoid direct exposure and mediate all communications through a single trusted individual (or very few of them). On the other hand, this particular aspect is not detectable in the unweighted analysis.   

\paragraph{Take-Home Message.} 
In short, our results confirmed the effectiveness of SNA in speeding up the process of disrupting criminal networks. Considering our datasets, we could severely affect network connectivity (with a 70\% drop) by neutralizing less than 5\% of the affiliates (either through sequential arrest of police raids). Betweeness centrality performed significantly better than the other three metrics, thanks to its specific focus on paths, rather than simple node degree. This is consistent with the typical operation of criminal networks where information diffuses through the shortest paths and within the organization, to minimize intra-affiliates interactions and, thus, the risk of interception. Therefore, law enforcement interventions should favor path-related centrality metrics (such as Betweenness) instead of other strategies.

\section*{Discussion}
\label{sec:discussion}

Our social capital investigation shows that the most significant centrality metric is Betwenness centrality. This is because it efficiently causes a deeper fragmentation in the Largest Connected Component (as visible in both networks under scrutiny). Betweeness is the only metrics (out the four analysed) that focuses on shortest paths, which reflects the {\em modus operandi} of ``cosca'' associations. Interfering on the paths produces a sensitive slow down in spreading messages among trusted affiliates. The same conclusion has been drawn regardless of weighting (to reflect interaction frequencies). 
Overall, we could substantially compromise connectivity by removing the top 5\% most influential nodes, computed according to Betweenness score.
This effect is achieved thanks to the network weights distribution (or rather, their concentration within few influential nodes), which leads to a rapid drop in $\rho$. Once the most influential nodes have been removed, the remaining ones are largely characterized by $w=1$, which makes the weighted and unweighted networks virtually indistinguishable. 
Our SNA results can be directly translated onto law enforcement actions, considering that we are now able to efficiently identify the top 5\% most  trusted affiliates (i.e., the ones typically employed as intermediaries between bosses and the other members). In turn, we can virtually neutralize the clans' internal communication infrastructure by getting the trusted affiliates in custody. Intuitively, whenever arrests can be made in block (raids), that would further impair the ability of the criminal communication network to be re-established. However, we have not studied this specific aspect, due to unavailability of necessary data.  

Generally speaking, the aim of a ``cosca'' is to conduct illegal activities, which would vary from place to place, is susceptible to local trends, and, ultimately pursues effective financial benefits. For instance, some clans may focus on drugs, rather than organ trafficking, prostitution, finance, or political influence. Quite commonly, clans pursue multiple activities, which makes it even more difficult to reconstruct the entangle of criminal communication networks (and perform SNA thereof).

Our datasets emerged directly from a set of official juridical acts, and focused on a single criminal activity (the securement of public procurement contracts): it involved a network of entrepreneurs; it was confined to a specific geographical area; and captured a definite time span. Thus, our dataset captured a relatively simpler snapshot of the complex entangle of mafia criminals, which constitutes both the strength and weakness of our study. 

On the one hand, the scope of our SNA is limited by the significance and breath of the datasets at hand. We have mentioned already how a more dynamic analysis of the network could not be done - for instance, to understand the re-connection ability following a disruption event. Also, we are capturing a single criminal activity in a confined spatio-temporal context. So, it was not possible to detect a broader and more diversified set of communications, such as those taking place in a more complex, multi-activity network. Nor could we detect external communications, such as those involving people who were not directly members of the criminal nets except for entrepreneurs (e.g., politicians, magistrates and businessmen).

On the other hand, the greater specificity of our networks allowed a cleaner analysis, focused on unveiling some hidden communication mechanisms. Having reduced the parameters under scrutiny and the complexity of the system, we could pinpoint a simple, yet effective direction to tackle criminal communication networks, which would have not necessarily emerged from a more complex network analysis. 
This simpler framework has allowed us to swiftly test out our hypothesis and to obtain reliable results, whereas larger networks would have not necessarily provided cleaner results, particularly when multiple criminal activities take place in parallel.

\section*{Conclusions}
\label{sec:conclusion}

In this work we focused on social capital analysis by studies on network disruption, using real-world data relating to a ``cosca'' that operated in Sicily (Italy) during the first decade of the 2000s. Our two datasets were derived from original juridical acts about two Sicilian clans who sought illegal profits from public procurement proceedings. 

To facilitate the replication and further extension of our work, we have placed an anonymized version of the datasets and the source code in a public repository  (\url{https://github.com/lcucav/networkdistruption}), including the \textit{Meeting} dataset (constructed from police stakeouts) and the \textit{Phone Calls} dataset (derived from police wiretaps). 

Starting from the datasets, we extracted both weighted and unweighted versions of the communications networks, presenting a preliminary characterization (weight distribution and shortest path length analysis) at a specialized conference ~\cite{10.1007/978-3-030-36683-4_36}.

In the present work, we have explored the mechanisms to identify key individuals in the network and, in turn, disrupt the information network through minimal node removal. We considered two disruption strategies, namely
\begin{enumerate*}[label=(\roman*)] 
\item a sequential node removal approach, and
\item block removal.
\end{enumerate*} 
The first one simulates the scenario in which the police arrest one ``cosca'' affiliate per time. The second one, mimics a police raid.
Next, we put to test four different centrality metrics, namely: 
\begin{enumerate*}[label=(\roman*)]
\item Degree centrality,
\item Betwenness centrality,
\item Katz centrality, and
\item  Collective Influence. 
\end{enumerate*} 
The effectiveness of the centrality metrics has been validated trough the $\rho$ parameter, which measures the drop of the Largest Connected Component size, after node removal, compared with the initial LCC size. 

The experiments unveiled  Betweenness as the most effective metric. Betweenness produced a greater impact in terms of network disruption, thanks to its prioritization of communication paths rather than individual nodes degree. Thus, the resulting winning strategy was to order nodes by Betweenness centrality score and proceeding with the removal of nodes starting from highest scores. This procedure tackles directly the communication mechanisms used in criminal networks, which are aimed at minimizing police interception.

This work is prone to considerable extensions and specializations. For one, it would be interesting to conduct a comparative study between social and human capital in Mafia associations. In fact, the resilience of criminal networks also depends on the personal qualities and competences of their members. In Network Science, those competences are represented as node labels that represent node roles. Therefore, in order to better assess the strength of a criminal organization, we should also look at the human capital endowment.

While we have looked at how to disrupt criminal networks by identifying the key information intermediaries, another promising angle is the identification of individuals holding highly specialized roles.   
Criminal organizations are increasingly infiltrating highly specialised activities that require very specific knowledge, skills and competence. 
For example, pharmacological and chemical knowledge are required in synthetic drug synthesis processes. The removal of these high human capital figures could lead to a further weakening of the resilience of criminal organizations, particularly because specialists are not always easily replaced.

\bibliographystyle{unsrt}
\bibliography{mybibfile}  

\end{document}